*Orphan High Field Superconductivity in Non-Superconducting Uranium Ditelluride*

Auth:

*Corey E. Frank,[1,2] Sylvia K, Lewin,[1,2] Gicela Saucedo Salas,[2,1] Peter Czajka,[1,2] Ian Hayes,[2] Hyeok Yoon,[2] Tristin Metz,[2] Johnpierre Paglione,[2] John Singleton,[3] *Nicholas P. Butch[1,2]

1. NIST Center for Neutron Research, National Institute of Standards and Technology, Gaithersburg, MD, USA
2. Maryland Quantum Materials Center, Department of Physics, University of Maryland, College Park, MD, USA
3. National High Magnetic Field Laboratory, Los Alamos National Laboratory, Los Alamos, NM, USA

*To whom correspondence should be addressed

**Abstract**

Reentrant superconductivity is a phenomenon in which the destructive effects of magnetic field on superconductivity are mitigated, allowing a zero-resistance state to survive under conditions that would otherwise destroy it. Typically, the reentrant superconducting region derives from a zero-field parent superconductor. Here, we show that in specifically-prepared $UTe_2$ crystals, extremely large magnetic field gives rise to an unprecedented high field superconductor that lacks a zero-field parent phase. This orphan superconductivity exists at fields between 37 T and 52 T, over a smaller angular range than observed in superconducting $UTe_2$. The stability of field-induced orphan superconductivity is a challenge to existing theoretical explanations, and underscores the likelihood of a field-induced modification of the electronic structure of $UTe_2$.

**Introduction**

Although uranium ditelluride ($UTe_2$) has been known since the mid-1900's to be a paramagnet,[1,2] its unconventional superconductivity was not reported until 2019.[3,4] In most superconductors, applied magnetic fields destroy superconductivity at and above the Pauli limit, the field at which Zeeman splitting destabilizes spin anti-aligned Cooper pairs.[5] In $UTe_2$, however, superconductivity survives to fields at least double the Pauli limit when the field is aligned in any crystallographic direction, indicating the presence of unconventional superconductivity.[3,4] When the applied field is aligned along the crystallographic *b* axis, superconductivity persists to a striking value of 35 T, above which it is sharply truncated by a metamagnetic transition into a field polarized state.[4] Moreover, superconductivity returns and persists up to an estimated 70 T at off-axis angles of 20º-40º from the crystallographic *b*- to *c*-axes.[6,7] This anisotropic and highly robust superconductivity strongly implies that $UTe_2$ is an intrinsic spin-triplet superconductor.[4,7] Owing to the potential for spin-triplet superconductors to host non-Abelian Majorana fermions,[8] such materials would be very attractive as building blocks for emergent technologies such as fault tolerant quantum computers.[9]

Despite intense research, the pairing state of the superconducting order parameter at low fields is yet to be unambiguously determined.[10-12] Determining the nature of the highest field superconductivity is even more of a challenge,[13-15] and the relationship between the low- and high-field superconducting phases is also uncertain. Although this question has been primarily addressed theoretically, there has been some experimental input. For example, NMR Knight-shift measurements suggest that there is a field-induced d-vector rotation involving a switch between $B_{3u}$ and $B_{2u}$ components.[11,16] Nonetheless, the most important effect underlying the intense field enhancement of superconductivity remains unclear. Leading explanations include lower dimensionality,[14,15] which can suppress the orbital limiting effects of magnetic fields, or internal exchange fields that counteract the applied external field,[4,7] leading the superconducting phase to experience smaller total magnetic fields.

To explore the relationship between the superconducting phases in UTe$_2$, we prepared samples of nominally non-superconducting UTe$_2$ via chemical vapor transport. The stability of superconductivity was mapped via magnetoresistance measurements that were performed in applied magnetic fields of up to 60 T with the field-angle rotated between the *b* and *c* axes. Our results show that these UTe$_2$ samples host an "orphaned" high field stabilized superconductivity without an accompanying low-field superconducting phase. In addition to being the field-stabilized superconductor, these findings dramatically limit possible explanations for the stability of high-field superconductivity in UTe$_2$, demanding a new theoretical framework.

**Results and Discussion**

UTe$_2$ crystalizes in a centrosymmetric orthorhombic structure (*Immm*, No. 71) with $D_{2h}$ point group symmetry. The anisotropy of this structure, coupled with the strong spin orbit coupling of 5*f* uranium, leads to strongly anisotropic response to applied magnetic field. Low-field superconductivity in UTe$_2$ persists to at least 8 T in all directions; superconductivity survives to the highest fields (~35 T) when *H* || *b*, the low-field magnetic hard axis. Whether the superconductivity along the b-axis consists of more than one phase is an open question.[12,17,18]

In the UTe$_2$ samples studied here, there is no evidence of superconductivity in any orientation for magnetic field smaller than 35 T. Instead, the samples are paramagnetic metals. Zero-field resistance measurements demonstrate Fermi-liquid $T^2$ dependence below 10 K (See Supplementary Information, Fig S.2) without evidence of a superconducting transition down to 110 mK, or 1/19 of the expected critical temperature,[3,4,7,19-23] reflecting the absence of zero-field superconductivity. One measure of the disorder in this sample is given by the residual resistivity ratio (RRR) of 7.5, which compares to a typical RRR of 18 – 30, as first discussed by Ran *et al*,[3] and recent high RRR (RRR ≈1000) grown via salt flux.[19]

Absent superconductivity, the dominant feature in the data (**Fig. 1**) is the metamagnetic transition at applied field, $H_m$. At $H_m$, the magnetization along the *b*-axis jumps discontinuously, and the system enters a field-polarized state. As shown in **Fig. 1**, the metamagnetic transition occurs just below 35 T along the *b* axis. This value is slightly lower than previous observations of $H_m$ reported from typically superconducting samples of UTe$_2$.[4,7,18,24-26]

The onset field of the metamagnetic transition in UTe$_2$ is of similar energy scale to the temperature at which there is a maximum in the magnetic susceptibility with field along *b*, $T_\chi^{max}$ ≈ 35 K, previously reported for both nonsuperconducting[27] and superconducting[6] UTe$_2$. This feature is typical of heavy fermion paramagnets with metamagnetic transitions, implying in those cases that Hm and $T_\chi^{max}$ are related by a single energy scale.[28] The agreement between the energy scales associated with $T_\chi^{max}$ and Hm is also important in UTe$_2$[24,25,29] and reflects the Kondo hybridization energy scale, as further observed in scanning tunneling microscopy[29] and magnetic excitations in inelastic neutron scattering experiments.[30] These results show that the heavy fermion state in UTe$_2$ is a robust characteristic.

We now consider the orphaned field-induced orphan superconducting phase (SC$_{FP}$) that occurs at fields greater than $H_m$ in the field polarized state. This SC$_{FP}$ phase, with boundaries defined as 50% of the observed transition, emerges close to a 29° offset from *b* to *c* and extends to 42° (**Fig. 1a**). The narrower angular range of the orphan SC$_{FP}$ is striking when compared to previously published data from typically superconducting UTe$_2$, which extends from 25° to 42°[4,24,25,31] (**Fig. S.6** in Supplementary Information). The orphan SC$_{FP}$ phase only survives to 52 T, compared to extrapolated values above 65 T in other reports.[4,7,32] Nevertheless, magnetoresistance (**Fig. 1b**) shows that the transition into the SC$_{FP}$ state is qualitatively similar to that in other samples. Note two important features: wider transitions as a function of field and a limited range of zero resistance, both as measured at 0.5 K. The zero-resistance state is centered at 36°, suggesting that there is no direct relationship between the stability of SC$_{FP}$ and the crystallographic (0 1 1) direction, situated at 23.7°.

The temperature dependence of orphan SC$_{FP}$ gives further information about the robustness of the superconductivity. The zero resistance state, measured at a 38.7° offset from *b* to *c*, persists to 0.5 K (**Fig. 2**), and a superconducting envelope persists to almost 0.9 K. All resistive signatures of superconductivity are suppressed by 1 K. This temperature differs dramatically from the value of 1.5 K reported before in samples exhibiting low field superconductivity.[4] Previously, the similar $T_c$'s of low-field and SC$_{FP}$ states lead to the inference that the two phases must involve similar pairing energies.[4] The presence of orphan SC$_{FP}$ suggests that this interpretation is incorrect, or that additional mechanisms must be considered, such as substantial differences in the effects of the same disorder on the smearing of the superconducting energy gaps, perhaps due to differences in gap structure.

The challenge of identifying any theoretical mechanism for field stabilization of SC$_{FP}$ is made more difficult by the absence of low-field superconductivity. Existing relevant theoretical attempts to describe high field superconductivity generally require the presence of zero-field superconductivity,[3,4,7,13-15,33]. It is instructive to review these mechanisms in light of the recontextualization demanded by the orphan $SC_{FP}$ phase. The magnetic field dependence of the superconductivity due to these mechanisms is illustrated in **Fig. 3**.

Reentrant superconductivity in organic superconductors and several chevrel phases is reportedly stabilized by the Jaccarino-Peter mechanism.[34-36] This mechanism involves an internal exchange field generated by the short-range magnetic fluctuations of localized moments which opposes the applied magnetic field, stabilizing reentrant superconductivity (**Fig. 3**).[37] For example, in the chevrel phase $Eu_{0.75}Sn_{0.25}Mo_6S_{7.2}Se_{0.8}$, zero-field superconductivity appears below 3.9 K and is suppressed by 1 T.[34] Above 4 T, the external field begins to adequately compensate for the internal exchange field, and superconductivity returns, persisting to approximately 22 T.[34] A similar mechanism is believed relevant to field-stabilized superconductivity in the antiferromagnetic insulator $\lambda$-$(BETS)_2FeCl_4$. Chemical substitution experiments show that the high field range of the superconductivity is decreased when antiferromagnetism is destabilized, and further indicate that $\lambda$-$(BETS)_2FeCl_4$ may have a "hidden" superconducting phase that competes with the antiferromagnetic internal field.[38]

However, it was pointed out previously that the Jaccarino-Peter mechanism is likely not appropriate for $UTe_2$[4] because this effect requires localized moments and is typically observed in experiment over a narrow angular field range.[37] This conclusion is reinforced by the new observations of orphan $SC_{FP}$. The absence of zero field superconductivity without magnetic order to generate a negative exchange field at $H = 0$ almost entirely precludes the compensation-effect as the primary field stabilizing mode in $UTe_2$.

Another possible explanation is that $SC_{FP}$ is stabilized by ferromagnetic fluctuations,[3] similar to field-reinforced superconductivity observed in ferromagnetic superconductors $UCoGe$[39] and $URhGe$[40] (**Fig 3**). In this model, stabilizing longitudinal spin fluctuations arise near a second-order ferromagnetic transition driven by magnetic field.[41] At ambient conditions, $UTe_2$ is also inferred to lie on the cusp of magnetic order, based on low field magnetometry at ambient[27] and high pressure[42]. However, an important caveat is that superconductors described by the spin-fluctuation model exhibit long range magnetic order in zero field, and show low-field superconductivity in addition to a magnetically reinforced superconducting phase.[39,40] More importantly, experiments show that the high-field phases in the ferromagnetic superconductors are *more readily suppressed* by temperature and disorder than the zero field phases,[39,40] so it is surprising to see the presumptive fragile phase without its more robust neighbor in $UTe_2$.

Another mechanism for stabilizing high field superconductivity involves field-induced Landau levels.[33] In this model, the field-induced orbits of conduction electrons are quantized, and eventually the increased cyclotron radii of quasiparticles orbiting the Fermi surface extends the coherence range of paired electrons, and thus the stability field of $H_{C2}$ (**Fig 3**). Hypothetically, superconductivity could be stabilized in this way at any temperature with sufficient field; however, typically the field strength required for this is far beyond the Pauli limit for spin-singlet superconductors.[33,43] Landau-level stabilization is most likely to be realized in low dimensional spin-triplet superconductors, and high pressure measurements of resistance in typical $UTe_2$ show phase transitions quantized with the signature $1/H$ relation.[44] A low-dimensional electronic structure may be inferred from angle-resolved photoemission spectroscopy,[45] and recent de Haas van Alphen oscillation measurements of low-field superconducting UTe$_2$ suggest quasi-two-dimensional cylindrical electron and hole Fermi surface sections.[46] However, the Fermi surface has three-dimensional characteristics [47,48] and whether the electronic structure has sufficiently low-dimensional character for relevant theories to apply remains an open question. Theoretical analysis has proposed that SC$_{FP}$ in UTe$_2$ is stabilized near the quantum limit by a Hofstadter butterfly regime of Landau level quantizations with large superlattices.[49] This stabilization regime implies the existence of an even higher field phase beyond SCFP, located at approximately 90 T.[44,49] Possible signatures of precursor effects related to Landau level stabilized superconductivity were reported in a previous high-field pressure study (ref 7). However, confirmation of this model would ideally involve observation of superconductivity in multiple Landau levels, requiring challenging measurements performed at significantly higher magnetic fields.

**Methods**

Single crystals of UTe$_2$ were grown as thin plates approximately 3 mm in length by chemical vapor transport with iodine as the transport agent. Approximately 1 g total of elemental U and Te in a 2:3 atomic ratio were sealed in an evacuated quartz ampule with 30 mg of iodine. The ampule was loaded into a two-zone horizontal furnace and the temperature was slowly increased to 800 ºC and 710 ºC in the charge and growth zones, respectively. Temperature was maintained for 1 week, after which transport was quenched by turning off power to the heating elements. Crystals grew as thin black plates in the *ab* plane (**Fig. S.1** in Supplementary Information). Crystallographic orientation was identified from the crystal habit.

Zero-field resistance measurements to 100 mK were performed on a Quantum Design Physical Property Measurement System (PPMS) using the adiabatic demagnetization refrigerator (ADR) option. Crystals were mounted on a cryogenic single axis goniometer,[50] and high field magnetoresistance measurements were performed at the National High Magnetic Field Laboratory (NHMFL), Los Alamos, NM using a 65 T short-pulse magnet. Identification of commercial equipment does not imply recommendation or endorsement by NIST.

**Funding**

This work was supported in part by the National Science Foundation under the Division of Materials Research Grant NSF-DMR 2105191. A portion of this work was performed at the National High Magnetic Field Laboratory (NHMFL), which is supported by National Science Foundation Cooperative Agreements DMR-1644779 and DMR-2128556, and the Department of Energy (DOE). JS acknowledges support from the DOE BES program "Science of 100 T", which permitted the design and construction of much of the specialized equipment used in the high-field studies. The authors declare no competing financial interest.

**Acknowledgements**

The authors would like to express their gratitude to University of Maryland undergraduate students: Patrick Chen, Elan Moskowitz, and Kimia Samieninejad, for their assistance in pulsed field measurement preparations.

**Author Contributions**

N.P.B. directed the project. C.E.F synthesized single crystalline samples. C.E.F., G.S.S., and S.K.L. oriented and prepared samples for magnetoresistance measurements. C.E.F, S.K.L, and J.S. performed the magnetoresistance measurements in the pulsed field. C.E.F. performed resistivity measurements at zero field. I.H., T.M., and H.Y. performed preliminary measurements. C.E.F. and N.P.B. wrote the manuscript with contributions from all authors.

**Figures**

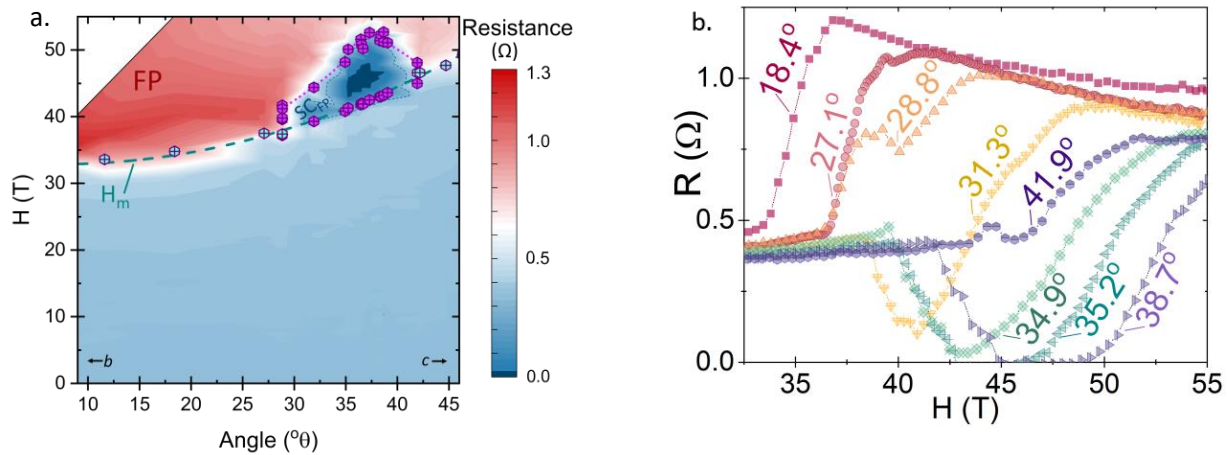

**Fig. 1 (a)** The angle dependence of the magnetoresistance (*b* to *c*, degrees) of orphan superconductivity in UTe$_2$ by applied field (H) at base temperature (approximately 0.5 K), with color indicating total resistance. Dark blue regions between 30-44° are where the sample resistance falls below the low field normal state value and the darkest color indicates zero resistance. Superconducting transitions (defined by 50% of the transition) are shown as purple hexagons, and transitions from the low field normal state to the field polarized normal state are indicated by blue hexagons. Lines are guides to the eye. **(b)** Magnetoresistance of orphan superconductivity at select angles near the SC$_{FP}$ phase (angles are in degrees from *b* to *c*). The large jumps in resistance near 35 T indicate the metamagnetic transition at applied fields H$_m$.

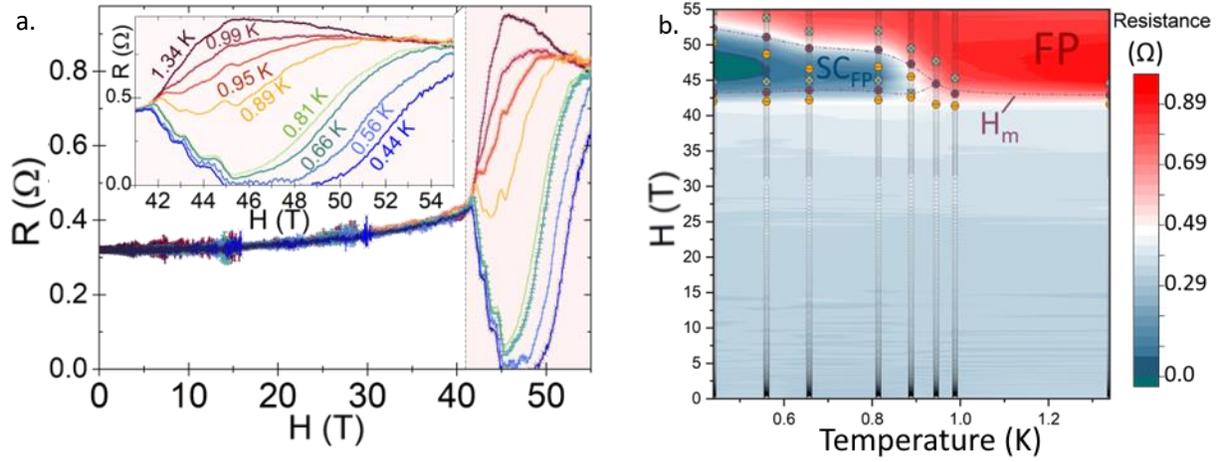

**Fig. 2** (**a**) The temperature dependence of the magnetoresistance of orphan superconductivity in UTe$_2$ at 38.7° between *b* and *c* versus applied field (H). Large jumps in resistance near 45 T indicate the metamagnetic transition. (**b**) The field-temperature phase diagram of orphan superconductivity. Open circles are features taken from data shown in a), and the R = 0 regions are highlighted in dark blue. Purple, orange barred, and light green crossed circles respectively indicate 50%, 10%, 90% of the transitions between superconducting and normal state or superconducting and field polarized state.

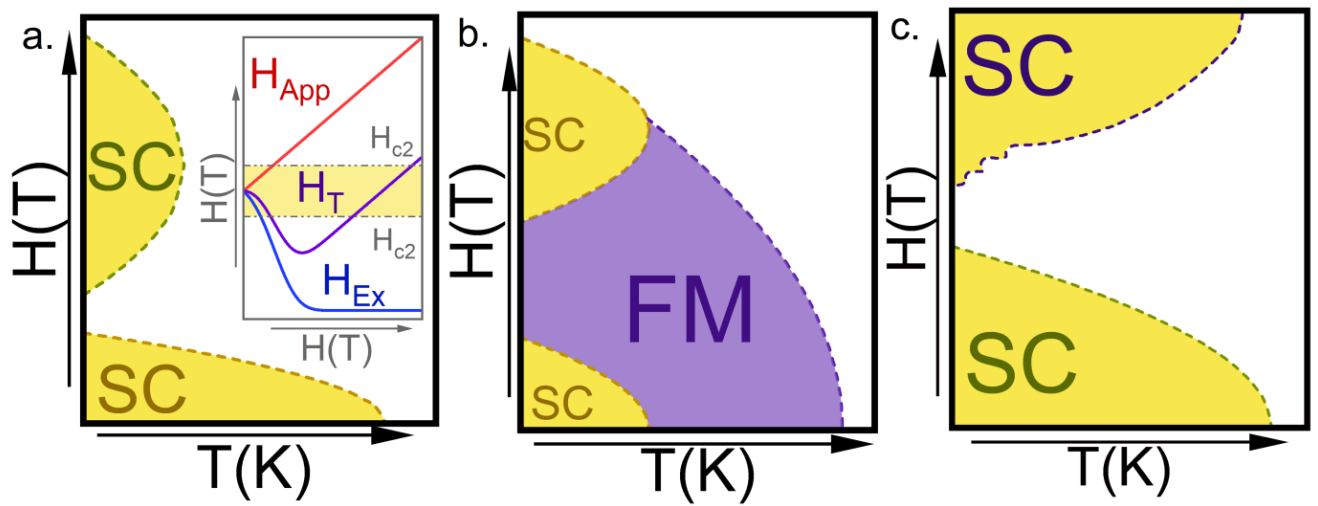

**Fig. 3.** Magnetic field – temperature schematic phase diagrams for superconductivity stabilized by different possible mechanisms. (3.a) The Jaccarino-Peter compensation effect. An internal exchange field (H$_{Ex}$, blue) opposes the applied field (H$_{App}$) resulting in reentrant superconductivity when the total internal field (H$_T$, purple) is smaller than H$_{c2}$. (3.b) Stabilization of ferromagnetic superconductivity near a quantum critical point. Strong magnetic fluctuations due to the destabilization of long-range magnetism enhance the superconducting pairing. Superconductivity can survive at and on either side of the QCP. (3c) Landau level stabilized superconductivity. The upper critical field of reentrant superconductivity is oscillatory in inverse field.